\documentclass[12pt]{JHEP}

\usepackage{amsmath}
\usepackage{amssymb,amsfonts}
\usepackage{latexsym}
\usepackage{mathtools}

\usepackage{enumerate}
\usepackage{empheq}
\usepackage{dsfont}
\usepackage{xcolor}
\let\normalcolor
\relax

\def\be{\begin{equation}}
\def\ee{\end{equation}}
\def\bea{\begin{eqnarray}}
\def\eea{\end{eqnarray}}

\def\G{{\sf G}}

\def\C{{\mathcal C}}

\newcommand{\Z}{\ensuremath{\mathbb Z}}
\newcommand{\R}{\ensuremath{\mathbb R}}

\newcommand{\abs}[1]{\left|#1\right|}

\usepackage{amsthm}

\renewcommand{\sp}{,~}

\title{Optimal Narain CFTs from Codes}

\author{Nikolaos Angelinos, Debarghya Chakraborty, Anatoly Dymarsky\\
	{Department of Physics and Astronomy, \\ University of Kentucky,\\Lexington, KY, USA 40506\\}}
\abstract{
	Recently established connection between additive codes and Narain CFTs provides a new tool to construct theories with special properties and solve modular bootstrap constraints by reducing them to algebraic identities. We generalize previous constructions to include many new theories, in particular we show that all known optimal Narain theories, i.e.~those maximizing the value of spectral gap, can be constructed from codes. For asymptotically large central  charge $c$ we show there are code theories with the spectral gap growing linearly with $c$,  with the coefficient 
	saturating the conjectural upper bound. We therefore conjecture that optimal Narain theories for any value of $c$ can be obtained from codes.
}

\def\pdfannotlink{\pdfstartlink}
\begin{document}

	\section{Introduction}
	Conformal modular bootstrap program aims to establish universal constraints on two-dimensional CFTs and elucidate properties of those special theories which saturate these constraints. One of the central goals of the modular bootstrap is to study theories maximizing the value of spectral gap for given fixed value of central charge \cite{Hellerman_2011}, as these theories for large central charge  are expected to be dual to weakly coupled gravity \cite{HKS}. To simplify this obviously challenging task one can restrict attention to a  class of Narain theories, i.e.~CFTs exhibiting $U(1)^c \times U(1)^c$ symmetry. In this case large spectral gap theories are not sparse (in the sense of \cite{HKS}), and their holographic description is less clear \cite{Dymarsky_2021}. Nevertheless study of such theories is well motivated  by both holography and the modular bootstrap, with the latter relating solutions of spinless bootstrap constraints to densest sphere packings \cite{hartman2019sphere}.\footnote{Here we are speaking of a  density of states satisfying (some subset) of modular bootstrap constraints, with no regard to whether there is an actual CFT yielding this density of states. Similarly, speaking of densest sphere packings, we in fact refer to a solution to Cohen-Elkies linear program constraints \cite{cohn2003new}, with no regard to whether there are actual associated sphere packings. }   Narain theories were studied in \cite{afkhami2020high} and \cite{afkhami2021free} using spinless and full modular bootstrap, with the hypothetical optimal theories being identified for $c\leq 8$.  
	Here, following \cite{nebe2006self} we say optimal to denote CFTs maximizing spectral gap for given $c$.
	
	A relation between quantum codes and Narain CFTs, proposed in \cite{dymarsky2021solutions}, generalizes chiral constructions of \cite{dolan1996conformal}. Starting from a code it constructs corresponding Narain lattice and expresses CFT torus partition function in terms of the code enumerator polynomial (for higher genus generalization see \cite{https://doi.org/10.48550/arxiv.2112.05168,https://doi.org/10.48550/arxiv.2205.00025}). In this way constraints of modular invariance reduce to two algebraic constrains at the level of enumerator polynomial.  The relation to quantum codes was recently extended and interpreted in terms of CFT Hilbert space in \cite{buican2021quantum}. There are also ``bottom-up'' generalizations when the connection with codes is perceived as a tool  to solve modular bootstrap constraints and  construct interesting CFTs \cite{ds,dymarsky2021non}, also see \cite{yahagi2022narain,furuta2022relation} for the subsequent developments.  In this paper we use this approach and introduce the umbrella construction which generalizes and encompasses the constructions of \cite{ds,dymarsky2021non}. In particular we show that all (conjecturally) optimal Narain theories for $c\leq 8$ identified in \cite{afkhami2021free} are in fact code theories, by providing an explicit way to construct these theories from codes. 
	
	Our construction should be understood as an infinite family of similar but distinct constructions. We consider codes over abelian groups  $\G=\Z_p \times \Z_q$ equipped with particular scalar product and map even self-dual codes $\C \subset \G^c$ to Narain lattices using a suitable generalization of the Construction A of  \cite{SPLAG}. The same group $\G$ might be mapped into lattices in several different ways, each way defining a particular construction. Then an appropriate generalization of the code Hamming distance (we explain what this is latter in the text), modulo certain subtleties, defines CFT spectral gap $\Delta^*$ such that ``better" codes with larger Hamming distance corresponds to larger $\Delta^*$.  In each case, the resulting Narain lattice necessarily has vectors of particular length which is independent of $c$. Hence any given construction can only yield CFTs with bounded spectral gap that doesn't grow with $c$. Nevertheless by considering a sequence of constructions parametrized by $c$ one can obtain a family of Narain theories with the spectral gap growing linearly with $c$. For $c\gg 1$ finding optimal codes, i.e. those maximizing corresponding Hamming distance,  is a challenging task, but one can average over a family of codes with the given $c$. From here we find that random code CFT, drawn from a particular ensemble, has spectral gap 
	\bea
	\Delta^*={c\over 2\pi e},\qquad c\rightarrow \infty,
	\eea
	which was conjectured in \cite{Dymarsky_2021} to  be asymptotically  largest possible value. Thus, we conclude that certain code CFT are optimal for $c\gg 1$ or at least give spectral gap with the conjectured maximal asymptotic value of $\Delta^*/c$. 
	
	The paper is organized as follows. In section \ref{construction} we outline our main construction mapping codes to Narain CFTs and then express their partition functions in terms of enumerator polynomials in section \ref{partition}. We then use these results to construct optimal  theories for $c\leq 8$  in section \ref{examples}. We  proceed by considering the case of asymptotically large $c$ and a family of associated constructions in section \ref{largec}. We conclude in section \ref{conclusions}.
	
	Note: recently a paper \cite{yahagi2022narain} appeared which has an overlap with the square lattice construction discussed in sections \ref{sglc}, \ref{psglc}.

	
	\section{Additive codes and Lorentzian lattices}
	\label{construction}
	The main ingredient of our construction is a 2-dimensional even  lattice $\Lambda \in \R^{1,1}$, which we call a glue lattice. Starting from such a lattice, we define an additive group $\G$, which serves as the alphabet of the code. Then standard Construction A maps a code $\C \subset \G^c$ into a lattice
	\bea
	\underbrace{\Lambda \oplus \dots \oplus \Lambda}_{c\, {\rm times}} \subset \Lambda_\C \subset \underbrace{\Lambda^\perp \oplus \dots \Lambda^\perp}_{c\, {\rm times}} \subset \R^{c,c}.
	\eea
	When $\C$ satisfies additional conditions, the lattice $\Lambda_\C$ is even and self-dual, thus defining a  Narain theory.

	\subsection{Even lattices in $\mathbb R^{1,1}$}
	We equip $\mathbb R^2$ with a Lorentzian metric 
	\be g=\begin{pmatrix}
		0 & 1\\
		1 & 0
	\end{pmatrix},\label{ip}\ee
	thus turning it into $\R^{1,1}$. 
	A two-dimensional ``glue''  lattice $\Lambda \subset \R^{1,1}$ is called integral if $v^Tgu\in\mathbb Z$ for any  $u,v\in\Lambda$. A lattice called even if $v^Tgv\in2\mathbb Z$ for any $v\in \Lambda$. Any even lattice is automatically integral.
	
	It is convenient to parametrize a  lattice by a generating matrix $\Lambda$, such that $v=\Lambda n$, $n\in \Z^2$ generates all lattice vectors. Here we abuse the notations by using $\Lambda$ to denote both the lattice and its generating matrix. The generating matrix is not unique, obviously $\Lambda$ and $\Lambda S$ for $S\in SL(2,\Z)$  generate the same lattice. So far we are only interested in the scalar product defined by \eqref{ip},
	we can identify all lattices related by $O(1,1)$, $\Lambda \sim O \Lambda$ for $O\in O(1,1)$. 
	
	By dual lattice $\Lambda^\perp$ we understand all vectors $v\in \Lambda^\perp\subset \R^{1,1}$ such that $v^Tgu\in\mathbb Z$ for any vector $u\in\Lambda$. At the level of generating matrix $\Lambda^\perp=g(\Lambda^{-1})^T$.
	
	We can parametrize all even lattices in $\R^{1,1}$ as follows. In full generality 
	\be 
	g_\Lambda=\Lambda^Tg\Lambda=\begin{pmatrix}
		2m & k\\
		k & 2n
	\end{pmatrix}\sp n,m,k\in\mathbb Z,\label{qf}
	\ee
	assuming $\det(\Lambda)=\sqrt{k^2-4mn}>0$.
	Then using $SL(2,\Z)$ we can bring $n,m,k$ to satisfy (see chapter 15 of \cite{SPLAG})
	\bea
	\label{splagcondition}
	0<k<\sqrt{k^2-4nm}<\min(k+2|n|,k+2|m|),
	\eea
	unless $k^2-4nm$ is a full square, in which case one can choose new $n,m,k$ such that 
	\bea
	n=0,\quad  -k<m\leq k. \label{splagcondition2}
	\eea
	
	With the help of an appropriate $O(1,1)$ transformation the corresponding generating matrix can be brought to the form 
	\bea \Lambda=\begin{pmatrix}
		1 & {n\over a}\\
		m & a
	\end{pmatrix}\sp a={1\over 2}(k+\sqrt{k^2-4mn}).
	\eea
	Since $\Lambda$ is integral, it is contained in its dual  $\Lambda^\perp\supseteq\Lambda$.
	We call the following quotient the ``glue group''
	\be 
	\label{defG}
	\G\:=\Lambda^\perp/\Lambda=\Z^2/g_\Lambda=\Z_p \times \Z_q. \ee
	Here  
	\bea
	p=gcd(2n,2m,k),\qquad q=|\G|/p, \label{p}
	\eea
	where $|{\sf G}|$ is the order of the group. This follows from the invariant factor decomposition of finitely generated abelian groups. 
	
	\be |\G|=|\det(\Lambda)/\det(\Lambda^\perp)|=k^2-4mn.\ee
	One possible parametrization of the elements of $\G$ is as pairs ${\sf g}=(a,b)$ of integer numbers $0\leq a<p,\, 0\leq b<q$. Another useful parametrization is as integer vectors $\ell({\sf g}) \in \Z^2$ modulo columns of $g_\Lambda$. An explicit map between these two parametrizations may be nontrivial.
	
	
	The metric on $\R^{1,1}$ induces a scalar product on $\G$, which we denote by $\eta$. It is defined up to shifts by integer numbers, which reflects different choices of representatives in the quotient $\Lambda^\perp/\Lambda$.
	One can consider codes over $\G$, namely additive linear subspaces $\C \subset \G^c$. A code would be called even if the norm of each element defined with $\eta$ is even
	\bea
	\eta({\sf c},{\sf c})=\sum_{i=1}^c \eta({\sf g}_i,{\sf g}_i)=\sum_{i=1}^c \ell({\sf g}_i)^T g_\Lambda^{-1} \ell({\sf g_i}) \in 2\Z,\quad {\sf c}\equiv ({\sf g}_1,\dots,{\sf g}_c)\in \C.
	\eea
	Despite that $\eta$ is  defined only  up to certain integer shifts, whether  $\eta({\sf g},{\sf g})$ is even does not depend on the representative of $\sf g$. 
	With help of $\eta$ one can also define dual code $\C^\perp$, as the group of all elements   ${\sf c}_1\in \C^\perp\subset \G^c$ satisfying $\eta({\sf c}_1,{\sf c}_2)\in \Z$ for any ${\sf c}_2\in \C$. A code is called self-dual when $\C=\C^\perp$.
	
	\subsection{Construction A}
	Starting from a  code $\C\subset \G^c$, Construction A associates to it the lattice  $\Lambda_\C$,
	\bea
	\underbrace{\Lambda \oplus \dots \oplus \Lambda}_{c\, {\rm times}} \subset \Lambda_\C \subset \underbrace{\Lambda^\perp \oplus \dots\oplus \Lambda^\perp}_{c\, {\rm times}} \subset \R^{c,c},
	\eea
	defined as the set of vectors in $\Lambda^\perp \oplus \dots \oplus\Lambda^\perp$ mapped into $\C\subset \G^c$ under \eqref{defG}. Then it is straightforward to see that an even $\C$ would give rise to an even lattice $\Lambda_\C$ and a self-dual $\C$ to a self-dual $\Lambda_\C$, both understood with respect to Lorentzian scalar product $g_L=g\oplus\dots \oplus g$ in $\R^{c,c}$.
	
	To define a Narain theory, besides Lorentzian scalar product,  even self-dual lattice should also be equipped with the Euclidean scalar product. For each $\Lambda$  defined in previous section there is $O(1,1)$ ambiguity how it can be embedded in $\R^2$. Thus, very explicitly we can write 
	\bea
	\label{LambdaC}
	v=\begin{pmatrix}
		O(\Lambda^\perp \ell_1+\Lambda k_1)\\
		\vdots\\
		O(\Lambda^\perp \ell_c+\Lambda k_c)\\
	\end{pmatrix}\in \Lambda_\C,\quad 
	(\ell_1,\dots, \ell_c)\in \C,\quad  k_i\in\mathbb Z^2, \label{ell}
	\eea
	where in \eqref{ell} we parametrize elements of $\G=\Z^2/g_\Lambda$ by vectors $\ell$. 
	Matrix $O$ is an arbitrary element from $O(1,1)$. In principle we can introduce $c$ different transformations $O_i\in O(1,1)$ acting in each $\R^{2}$ plane. In this case most of the construction remains intact, but the permutation of  factors of $\G$ in  $\C\subset \G^c$, which is conventionally considered to be a code equivalence, would no longer yield physically equivalent lattices. In what follows  we assume that all factors $O$ are the same. 
	
	The main result of this section is as follows. For any glue group $\G$ defined via \eqref{defG} with the help of an appropriate even lattice $\Lambda \subset \R^{1,1}$, any even self-dual code $\C\subset \G^c$ via Construction A defines Narain lattice $\Lambda_\C$ \eqref{LambdaC} and hence a Narain CFT. We will call such CFTs code theories.

	\subsection{Example: square glue lattice}
	\label{sglc}
	Consider the following glue lattice  generating matrix 
	\be \label{sgl}
	\Lambda=\sqrt{p}\,g,\quad p\in\mathbb N.\ee
	The dual lattice is generated by $\Lambda^\perp=I/\sqrt{p}$. Clearly this is the case of $n=m=0,k=p$ and the glue group $\G=\Z_p\times \Z_p$ is parametrized by ${\sf g}=(a,b)\in \G$, $0\leq a,b<p$ and 
	\bea
	\ell({\sf g})=(a,b)^T.
	\eea
	
	It is convenient to write codewords ${\sf c}=({\sf g_1},\dots, {\sf g}_c)\in \C\subset \G^c$ as
	\bea
	{\sf c}=(a_1,\dots,a_c|b_1,\dots,b_c)\in \Z_p^{2c}.
	\eea
	A code can be defined with a $2c\times d$ generating matrix $G_\C$ such that
	\bea
	{\sf c}=G_\C\, r,\quad r\in \Z_p^{d},
	\eea
	where $d$ depends on $p$. For prime $p$ generating matrix is $2c \times c$ and using permutations can always be brought to the form 
	\bea
	\label{Bcodes}
	G_\C=\left(I\,|\,B^T\right),
	\eea
	where $B$ is an integer values antisymmetric $c\times c$ matrix defined mod  $p$,
	\bea
	B^T=-B\mod\, p,
	\eea and $B_{ii}=0$. Applying Construction A to such codes leads to Narain lattices generated by 
	\bea
	\Lambda_\C=\left(\begin{array}{cc}
		p I_c & B\\
		0 & I_c
	\end{array}\right)/\sqrt{p},\qquad g_L=\left(\begin{array}{cc}
		0 & I_c \\
		I_c & 0
	\end{array}\right), \label{B}
	\eea
	associated with the Lorentzian scalar product $g_L$. Here $I_c$ is a $c$-dimensional identity matrix. 
	
	The form of $\Lambda_\C$ provides a clear interpretation -- corresponding Narain theory describes $c$ scalars compactified on a $c$-dimensional cube of size $1/\sqrt{p}$ in presence of B-field $B$. 
	
	The construction described above is a straightforward generalization of the original construction of \cite{ds,dymarsky2021solutions}, which considered the case of $p=2$. It has been recently  introduced independently in \cite{yahagi2022narain}.

	\subsection{Generalization: isodual codes}
	Permutation of factors $S:\G^c \rightarrow \G^c$
	\bea
	S: ({\sf g}_1,\dots ,{\sf g}_c) \rightarrow ({\sf g}_{i_1},\dots ,{\sf g}_{i_c})
	\eea
	is the simplest example of code equivalences, defined as a linear transformation 
	$S:\G^c \rightarrow \G^c$ which preserves scalar product $\eta$.
	Provided dual code is equivalent to the original one 
	\bea
	\C^\perp=S(\C), \label{isodualS}
	\eea
	such a code is called isodual. From this follows $S^2=1$, i.e.~when $S$ is a permutation, it is a pair-wise permutation, with the corresponding matrix  satisfying $S^T=S$. 
	
	We can introduce evenness as the condition for all codeword $({\sf g}_1,\dots, {\sf g}_c)\in \C\subset \G^c$ to have even scalar product with its permuted self, 
	\bea
	\label{evennesS}
	\sum_{i=1}^c \eta ({\sf g}_i,{\sf g}_{S(i)})\in 2\Z.
	\eea
	
	An even, isodual code $\C$ with respect to some pairwise permutation $S$, via Construction  A \eqref{LambdaC} gives rise to an even lattice, which is self-dual with respect to Lorentzian scalar product
	\bea
	g_L=g\otimes S. \label{gL}
	\eea
	In this way isodual codes also can be used to define Narain lattices and code CFTs.

	\subsection{CFT spectral gap and code modified Hamming distance}
	The capacity of a code to preserve information is rooted in the ability to unambiguously restore ${\sf c}\in \C$ in case it got corrupted during transmission process. Speaking colloquially, the code is better (stronger), provided all codewords are  maximally distinct from each other. To quantify that coding theory uses Hamming distance $d({\sf c},{\sf c}')$, defined as the number of different components of ${\sf c}$, ${\sf c}'$. For the binary linear code, when the components of ${\sf c}$ are either zero or one, $
	d({\sf c},{\sf c}')=|{\sf c}-{\sf c}'|_1=|{\sf c}-{\sf c}'|_2^2$. When the code is not binary, there could be various generalizations of Hamming distance, relevant for different practical scenarios. Thus, if the transmission error changes a codeword component's value randomly, with the new value being unrelated to the original one, then Hamming distance defined as the number of distinct components is still relevant. Yet, in many practical settings the nature of errors are such that  the new values tend to be closer to  old ones, which suggest using various metrics. In particular in case of codes over $\Z_p$ one often defines 
	\bea
	\label{norm}
	d({\sf c},{\sf c}')=|{\sf c}-{\sf c}'|_2^2 =\sum_{i=1}^{c} |a_i-a'_i|^2,\\
	{\sf c}=(a_1,\dots,a_c)\in \C\subset \Z_p^c,\quad 0\leq a_i<p.
	\eea
	For a linear code, the code distance is 
	\bea
	d(\C)=\min_{0\leq i,j\leq |\C|-1}^{} d({\sf c}_i,{\sf c}_j)=\min_{{\sf c}\in \C,{\sf c}\neq 0} d(0,{\sf c}).
	\eea
	The value $w({\sf c}):=d(0,{\sf c})$ is usually called the weight of ${\sf c}$.
	
	Construction A \eqref{LambdaC} relates each codeword ${\sf c}\in \C$ to a family of vectors 
	$v({\sf c})$. We define the weight of ${\sf c}$ to be the minimal length square among all such $v$, 
	\bea
	w({\sf c})&=&\min_{k_1,\dots, k_c\in \Z} |v|^2,\\
	v&=&(\ell_1+\Lambda k_1,\dots,\ell_c+\Lambda k_c),\qquad {\sf c}=(\ell_1,\dots, \ell_c)\in \C\subset \G^c,\quad \G=\Z^2/g_\Lambda. \nonumber
	\eea
	The weight $w({\sf d})$ is closely related to the norm of ${\sf c}$ calculated with the scalar product $\eta$ inherited from $\R^{1,1}$ on $\G$. 
	
	Clearly, from this definition follows that code theory spectral gap $\Delta^*$, defined as the length-squared of the shortest non-zero vector divided by two, is simply related to code's generalized Hamming distance
	$\Delta^*={d(\C)\over 2}.$
	This relation is transparent, but there is one caveat: zero codeword ${\sf c}=0$ is mapped into the origin of $\Lambda_\C$, as well as many vectors of the form $\Lambda(k_1,\dots, k_c)$. The origin is excluded from the consideration, while minimizing over $k_i$ yields shortest vector of $\Lambda$. We thus have 
	\bea
	\label{sgf}
	\Delta^*={1\over 2}\min\left({d(\C)},{|v_\Lambda|^2}\right),
	\eea
	where by $|v_\Lambda|$  we denoted the length of shortest non-trivial vector of $\Lambda$.  This length depends non-trivially on the choice of $n,m,k$ and the $O(1,1)$ factor (which we absorbed into the definition of $\Lambda$), but an upper bound \eqref{2dbound} is readily available, see below.
	
	For the square lattice of subsection \ref{sglc} we find that $d({\sf c},{\sf c'})$ is given by \eqref{norm}
	where by $|a-a'|^2$ with $0\leq a,a'  <p$ we understand 
	\bea
	|a-a'|^2:=\min_{k\in Z}\, (a-a'+p k)^2,
	\eea
	and $|v_\Lambda|^2=p$.
	
	To obtain the upper bound on the length, in the sense of Euclidean norm, of the shortest  vector $v_\Lambda$ belonging to   
	\eqref{LambdaC} we  consider all $\ell_i=0$, arbitrary $k_1$ and $k_i=0$ for $i>1$. Then the Euclidean norm of corresponding two-dimensional vectors is
	\bea
	|v|^2=k_1^T \Lambda^T O^T O \Lambda k_1, \label{2dmetric}
	\eea
	which defines a positive-definite scalar product in $\R^2$. The shortest vector will necessarily be shorter than $|v_\Lambda|^2\leq 2|\G|^{1/2}/\sqrt{3}$, see Appendix \ref{Ebound}, and therefore corresponding code CFT would have the spectral gap not exceeding 
	\bea
	\Delta^*\leq {\sqrt{k^2-4 nm}\over \sqrt{3}}. \label{2dbound}
	\eea
	This is a standard weakness of the Construction A lattices; they always include short vectors of a certain length, which does not increase with $c$. Therefore to construct large spectral gap CFTs with $\Delta^*$ scaling linearly with $c$ we would need to consider a cascade of different constructions by adjusting $k,n,m$ together with $c$. 
	
	\section{Torus partition function of code theories }
	\label{partition}
	\subsection{Enumerator polynomial and theta-series}
	One of the central properties of code theories which make them interesting is that their torus partition function can be expressed in a compact way in terms of  the so-called code enumerator polynomial which characterizes the corresponding code. Generalization of this result to higher genus partition function is also possible \cite{henriksson2021codes,henriksson2022narain}.
	
	We first define the full enumerator polynomial of a code $\C$  as a vehicle to  count how many times each element ${\sf g}\in \G$ appears in each codeword of $\C$,
	\be 
	P_C(\{x_{\sf g}\})=\sum_{({\sf g}_1,\dots,{\sf g}_c)\in \C} \prod_{i=1}^c x_{{\sf g}_i}.
	\ee
	This is a degree $n$ homogeneous polynomial of $|\G|$ variables $x_{\sf g}$. 
	
	The polynomial of the dual code $\C^\perp$ is related to the polynomial of $\C$ by the MacWilliams identity
	\be  \label{SD}
	P_{\C^\perp}(\{\tilde x_{\sf g}\})=P_\C(\{x_{\sf g}\}),\ee
	where
	\be \label{MW}
	\tilde x_{\tilde{\sf g}}= {1\over \sqrt{\abs{\G}}}\sum_{{\sf g}\in \G} \exp(-2\pi i \eta(\tilde{\sf g},{\sf g}))x_{\sf g}.\ee
	In particular, this means that the  enumerator polynomial of a self-dual code is invariant under the above transformation.
	
	The torus partition function of  a code CFT, associated with $\Lambda_\C$ obtained by Construction A, is given in terms of $P_\C$, 
	\be 
	\mathcal Z_{\C}(\tau)={P_\C(\{\psi_{\sf g}(\tau)\})},\label{partfunc}
	\ee
	where
	\bea
	\label{psi}
	\psi_\ell(\tau)={1\over |\eta(\tau)|^2}\sum_{k\in\mathbb Z^2}\exp(i \pi v^T\Omega v),\\
	v=\Lambda^\perp \ell+\Lambda\, k,\quad  
	\Omega=\begin{pmatrix}
		i\tau_2 & \tau_1\\
		\tau_1 & i\tau_2
	\end{pmatrix}, 
	\eea
	$\tau=\tau_1+i\tau_2$ is the torus modular parameter and in \eqref{psi} we parametrize elements  of $\G$ with help of vectors $\ell \in \Z^2/g_\Lambda$. We also absorbed $O\in O(1,1)$ into the definition of $\Lambda$.  
	
	The modular group, generated by $T:\tau\to\tau+1$ and $S:\tau\to-1/\tau$, transforms  $\Omega$ as follows
	\be T\circ\Omega=\Omega+g,\ee
	\be S\circ\Omega=-\Omega^{-1}.\ee
	Functions $\psi_\ell$ transform accordingly 
	\bea \label{TT}
	T\circ\psi_\ell(\tau)&=&\psi_\ell(\tau+1)=\exp(i\pi v^T g _\Lambda v )\psi_\ell(\tau),\quad v=\Lambda^\perp \ell,\\
	\label{SS}
	S\circ\psi_\ell(\tau)&=&\psi_\ell(-1/\tau)={1\over \sqrt{\abs{\G}}}\sum_{\ell'\in \G=\Z^2/g_\Lambda} \exp(-2\pi i u^T g_\Lambda v)\psi_{\ell'}(\tau),\quad u=\Lambda^\perp \ell'.\qquad
	\eea
	Evenness and self-duality of $\C$  ensure that $\mathcal Z_{\C}(\tau)$ is invariant under $T$ and $S$ respectively. Indeed, since the code is even, for any $(\ell_1,\dots,\ell_n)\in \C$ we have $\sum_i \ell_i^T g_\Lambda^{-1} \ell_i\in 2\Z$, and  therefore $T$ is a symmetry of  $P_\C(\{\psi_\ell\})$, while it is invariant under \eqref{SS} because of self-duality (\ref{SD},\ref{MW}). 
	
	\subsection{Example: theta series for square glue lattice}
	\label{psglc}
	For the lattice \eqref{sgl} functions $\psi_{\sf g}$ defined in (\ref{psi}) read
	\be \psi_{ab}={1\over |\eta|^2}\sum_{k_1,k_2\in\Z} q^{{p\over 4}({a+b\over p}+k_1+k_2)^2}\bar q^{{p\over 4}({a-b\over p}+k_1-k_2)^2},\ee
	where $(a,b)\in \G=\mathbb Z_p\times\mathbb Z_p$.
	This can be written as follows
	\be 
	|\eta|^2\psi_{ab}=\Theta_{a+b,p}\bar\Theta_{a-b,p}+\Theta_{a+b-p,p}\bar\Theta_{a-b-p,p},
	\ee
	where
	\be 
	\Theta_{m,p}=\sum_{n\in \Z} q^{p(n+{m\over 2p})^2}.
	\ee
	These functions are the chiral algebra characters  of  free boson compactified at radius $R=\sqrt{2k}$.

	Note, if we perform $O(1,1)$ rotation on the lattice $\Lambda$, functions $\psi_{ab}$ will change. Let
	\be \Lambda'=O\sqrt pg \sp O=\begin{pmatrix}
		\lambda &0\\
		0 & \lambda^{-1}
	\end{pmatrix}.\ee
	Then
	\be 
	|\eta|^2\psi_{ab}=\sum_{k_1,k_2\in \Z} q^{{p\over 4}({\lambda a+\lambda^{-1}b\over p}+\lambda k_1+\lambda^{-1}k_2)^2}\bar q^{{p\over 4}({\lambda a-\lambda^{-1}b\over p}+\lambda k_1-\lambda^{-1}k_2)^2} .\ee
	For $\lambda=\sqrt{q}$ with $q\in\mathbb N$, we can again decompose $\psi_{ab}$ as follows
	\be 
	|\eta|^2\psi_{ab}=\sum_{k=0}^{q-1}\Theta_{q a+b+kp,qp}\bar\Theta_{qa-b-kp,qp}+\Theta_{q(a-p)+b+kp,qp}\bar\Theta_{q(c_1-p)-c_2-kp,qp},\label{id1}\ee
	where the functions $\Theta_{m,pq}$ above are now characters of compactified boson at radius $R=\sqrt{2pq}$.
	
	Finally, if $\lambda=\sqrt{q/r}$ with $q,r$ are co-prime, we can again perform the same decomposition to obtain a more general result
	\be\begin{split} 
		|\eta|^2\psi_{ab}=\sum_{v_1=0}^{r-1}\sum_{v_2=0}^{q-1}&\Theta_{q(a+pv_1)+r(b+pv_2),pqr}\bar\Theta_{q(a+pv_1)-r(b+pv_2),pqr}+\\&\Theta_{q(a+pv_1)+r(b+pv_2)-pqr,pqr}\bar\Theta_{q(a+pv_1)-r(b+pv_2)-pqr,pqr} ,\end{split}\ee
	where the functions $\Theta_{m,pqr}$ above are characters of compactified boson at radius $R=\sqrt{2pqr}$.

	\subsection{Partition function in case of isodual codes}
	In case of isodual codes satisfying \eqref{isodualS} with pairwise permutation $S$, 
	the codeword ${\sf c}=({\sf g}_1,\dots, {\sf g}_c)\in \C$ should be understood as consisting of $r$ pairs $({\sf g}_i,{\sf g}_j)$ with $S$ mapping $i \leftrightarrow j$, while the remaining $c-2r$ ``letters'' remain intact. It is convenient to introduce new notation for ${\sf c}$ which is related to the previous one by permutation,
	\bea
	{\sf c}=(({\sf g}_{i_1},{\sf g}_{j_1}),\dots ,({\sf g}_{i_{r}},{\sf g}_{j_{r}}),{\sf g}_{2r+1},\dots , {\sf g}_c)\in \C.
	\eea
	With this notation we define an extended enumerator polynomial, which will depend on both $\C$ and $S$. It is a function of  $|\G|^2$ variables  $y_{{\sf g}_1{\sf g}_2}$ 
and $|\G|$ variables $x_{\sf g}$,
	\bea
	P_\C^S(\{y_{{\sf g}_1{\sf g}_2}\},\{x_{\sf g}\})=\sum_{(({\sf g}_{i_1},{\sf g}_{j_1}),\dots ,({\sf g}_{i_{r}},{\sf g}_{j_{r}}),{\sf g}_{2r+1},\dots , {\sf g}_c)\in \C}\,  \prod_{k=1}^{r} y_{{\sf g}_{i_k}{\sf g}_{j_k}} \prod_{i=2r+1}^c x_{{\sf g}_i}.
	\eea
	The CFT partition function is given by 
	\be 
	\mathcal Z_{\C}(\tau)={P^S_\C(\{\psi_{{\sf g}_1{\sf g}_2}(\tau)\},\{\psi_{\sf g}(\tau)\})},\label{partfuncS}
	\ee
	where
	\bea
	\label{psiS}
	\psi_{\ell_1 \ell_2}(\tau)={1\over |\eta(\tau)|^4}\sum_{k_1,k_2\in\mathbb Z^2}\exp(i\pi(v_1,v_2)^T\tilde\Omega(v_1,v_2)),\\
	\tilde\Omega=\begin{pmatrix}
		i\tau_2 I_2 & \tau_1 g\\
		\tau_1 g & i\tau_2 I_2 
	\end{pmatrix},\qquad 
	v_i=\Lambda^\perp \ell_i+\Lambda\, k_i.
	\eea
	Under modular transformations $T:\tau\to\tau+1$ and $S:\tau\to-1/\tau$, this function changes as follows 
	\bea 
	T\circ\tilde\Omega&=&\tilde\Omega+g\otimes g,\\
	S\circ\tilde\Omega&=&-\tilde\Omega^{-1}.
	\eea
	and 
	\bea 
	\label{TSS} 
	T\circ\psi_{\ell_1 \ell_2}&=&\exp(2i\pi v_1^T g v_2)\psi_{\ell_1 \ell_2},\quad v_i=\Lambda^\perp \ell_i,\\
	S\circ\psi_{\ell_1 \ell_2}&=&{1\over |\G|} \sum_{\ell_1',\ell_2' \in \G=\Z^2/g_\Lambda}\exp(2i\pi (u_1^Tgv_1+u_2^Tg v_2))\psi_{\ell_1' \ell_2'},\quad  u_i=\Lambda^\perp \ell_i'. \ \ \  \label{SSS}
	\eea
	Clearly when $\C$ is even in the sense of \eqref{evennesS} and isodual in the sense of  \eqref{isodualS}, the identities (\ref{TSS},\ref{SSS}) respectively ensure modular invariance of \eqref{partfuncS}.

	\section{Examples: optimal Narain theories for small $c$}
	\label{examples}
	In this section we consider a number of explicit examples of code theories. In particular we  discuss  optimal theories, i.e.~those with the largest spectral gap, for $c\leq 8$ identified in \cite{afkhami2021free}, and show they all are codes theories, in the sense defined in this paper. 
	
	\subsection{$c=1$}
	We first consider the simplest case of $n=m=0$, when 
	\bea
	g_\Lambda=\left(\begin{array}{cc}
		0 & k \\
		k & 0 \end{array}\right).
	\eea
	In this case the group $\G=\Z_k \times \Z_k$ is parametrized by vectors  $\ell=(a,b)$ for $0\leq a,b<k$. Let's consider a self-dual code $\C=\C^\perp$ and demand it to be even and self-dual. When $k$ is prime the only such two codes consist or vectors $(a,0)$ and  $(0,a)$ for $0\leq a<k$ correspondingly. 
	Using appropriate  $O(1,1)$ transformation we can bring corresponding Narain lattice to the form 
	\bea
	\Lambda_\C\ni \left(\begin{array}{c} a/\sqrt{k} \\ b\sqrt{k}\end{array}\right),\qquad a,b\in \Z.
	\eea
	At this point we recognize Narain lattice of a compact boson of radius $R^2=2k$. Choosing $k=1$ will yield boson at self-dual radius, which has largest possible spectral gap 
	\bea
	\Delta^*={1\over 2}.
	\eea
	
	The corresponding enumerator polynomial is simply $P=x_{00}$, giving rise to  torus  partition function via \eqref{partfunc} and \eqref{psi},
	\bea
	\mathcal Z_{\mathcal C}(\tau,\bar\tau)=\Psi_{0,0}={1\over |\eta|^2}\sum_{n,m\in Z} q^{(m+n)^2\over 4}\bar q^{(m-n)^2\over 4}={\abs{\theta_3(2\tau)}^2+\abs{\theta_2(2\tau)}^2\over |\eta|^2}.
	\eea
	
	
	Clearly, an appropriate $O(1,1)$ transformation will turn $\Lambda_\C$ to any other Narain lattice in $\R^{1,1}$, or, equivalently, change the compact boson radius $R$ to any desired value. In other words, together with the $O(1,1)$ factor our code construction is versatile enough such that any $c=1$ Narain theory is a code theory. This emphasizes the bottom-up nature of our approach. While codes are expected to reflect some algebraic properties of the underlying CFTs in the top-down constructions \cite{buican2021quantum}, in our construction certain non-rational CFTs  without obvious algebraic properties which would make them ``finite'' also can be obtained from codes.

	\subsection{$c=2$}
	We start with $m=2,n=-1,k=2$, which satisfies \eqref{splagcondition} and the glue lattice generated by 
	\bea
	g_\Lambda=\Lambda^T g \Lambda=\left(
	\begin{array}{cc}
		4 & \ \  2 \\
		2 & -2 \\
	\end{array}
	\right),\quad 
	\Lambda=R\,\, 2\left(\begin{array}{cc}
		1 & 1/2 \\
		0 & \sqrt{3}/2 \end{array}\right),\quad R=\left(\begin{array}{cc}
		1 & -1 \\
		1 & \, \, 1\end{array}\right)/\sqrt{2}.
	\eea
	From the Euclidean point of view this is a hexagonal (triangular) lattice with the lattice vectors of length $2$, rotated by $\pi/2$. Using equivalence transformation 
	\bea
	P=\left(
	\begin{array}{cc}
		0 & 1 \\
		-1 & 1 \\
	\end{array}
	\right)\in SL(2,\Z)
	\eea
	we can bring $g_\Lambda=\Lambda^T g\Lambda$ to the diagonal form 
	\bea
	\left(
	\begin{array}{cc}
		-2 & 0 \\
		0 & 6 \\
	\end{array}
	\right)=P^T g_\Lambda P,
	\eea
	which makes decomposition $\G=\Z_2  \times \Z_6$ manifest, with the map
	\bea
	{\sf g}=(a,b)\in \G,\quad 0\leq a<2,0\leq b<6,\qquad \ell({\sf g})=(P^T)^{-1}\left(\begin{array}{c} a\\ b \end{array}\right)\in \Z^2/g_\Lambda.\nonumber 
	\eea
	
	We consider a code $\C$ generated by the following three codewords
	\bea
	{\sf c_1}=((0,3),(1,0)),\\
	{\sf c_2}=((1,0),(0,1)),\\
	{\sf c_3}=((0,0),(0,2)),
	\eea
	in the notations ${\sf c}=({\sf g}_1,{\sf g}_2)=((a_1,b_1),(a_2,b_2))$. This codes is iso-dual, $\C^\perp=S(\C)$, where $S$ is the permutation of two elements. Corresponding lattice $\Lambda_\C$ obtained via \eqref{LambdaC}
	\bea
	\Lambda_\C=\left(\begin{array}{cc}
		\Lambda^\perp (P^T)^{-1} & 0\\
		0 & \Lambda^\perp (P^T)^{-1}
	\end{array}\right)\left(\ell+\left(
	\begin{array}{c}
		2k_1\\
		6 k_2\\
		2k_3\\
		6 k_4\end{array}\right)\right),\quad \ell^T=\sum_{i=1}^3 n_i {\sf c}_i,\quad  n_i\in \Z,\quad k_i\in \Z. \nonumber
	\eea
	In this expression above we should understand codewords ${\sf c}_i$ as regular vectors in $\Z^4$. This lattice is a Narain lattice with respect to the Lorentzian metric \eqref{gL}
	\bea
	g_L=\left(\begin{array}{cc}
		0 & S\\
		S & 0
	\end{array}\right),\qquad S=
	\left(\begin{array}{cc}
		0 & 1 \\
		1 & 0
	\end{array}\right).
	\eea
	By an orthogonal transformation $g_L$ can be brought to conventional form 
	\bea
	O g_L O^T=
	\left(\begin{array}{cc}
		0 & I_2\\
		I_2 & 0
	\end{array}\right),\qquad O={1\over\sqrt2} \left(
	\begin{array}{cccc}
		1 & 1 & 0 & 0 \\
		0 & 0 & 1 & -1 \\
		0 & 0 & 1 & 1 \\
		-1 & 1 & 0 & 0 \\
	\end{array}
	\right),
	\eea 
	such that $\Lambda_\C$ becomes equivalent to the Narain lattice $\Lambda_{c=2}$ defining $SU(3)_1$ WZW theory
	\bea
	\nonumber
	\Lambda_{c=2}&\sim& O\Lambda_C,\quad 
	\Lambda_{c=2}=\left(\begin{array}{cc}
		(\gamma^{-1})^T & B \gamma \\
		0 & \gamma \end{array}\right),\quad 
	\gamma=
	\sqrt{b_2\over t_2}
	\begin{pmatrix}
		1 & t_1\\
		0 & t_2
	\end{pmatrix},\qquad B={b_1\over b_2}\begin{pmatrix}
		0 & 1\\
		-1 & 0
	\end{pmatrix},
	\eea
	where $t_1+i t_2=b_1+i b_2=(1+i\sqrt{3})/2$.
	
	The code enumerator polynomial of $\C$ is
	\bea
	\nonumber
	P^S_\C=y_{00,00}+y_{00,02}+y_{00,04}+y_{10,01}+y_{10,03}+y_{10,05}+y_{03,10}+y_{03,12}+y_{03,14}+y_{13,11}+y_{13,13}+y_{13,15},
	\eea
	which yields partition function via \eqref{partfuncS}.
	Shortest lattice vector with $\ell^T={\sf c}_2$ or $\ell^T={\sf c}_3$ and $k_i=0$ has length $|v|^2=4/3$, hence corresponding CFT has spectral gap $\Delta^*=2/3$.

	\subsection{$c=3,4,5$}
	Optimal theories for $c=3,4,5$ were constructed from codes in \cite{ds}. They correspond to $k=2$ and $n,m=0$ with 
	\bea
	\Lambda=\sqrt{2}\left(\begin{array}{cc}
		0 & 1\\
		1 & 0
	\end{array}\right),\qquad g_\Lambda=2 \left(\begin{array}{cc}
		0 & 1\\
		1 & 0
	\end{array}\right),\qquad \Lambda^\perp = I_2/\sqrt{2}.
	\eea In this case $\G=\Z_2\times \Z_2$ which as an additive group is equivalent to $F_4$. 
	As discussed in section \ref{sglc} these codes are parametrized by B-matrix controlling the B-field of the Narain compactification, $B^T=B \mod 2$, see \eqref{B}. The case of $k=2$ is special because  antisymmetric $B^T=-B \mod 2$ and
	symmetric matrices $B$ are equivalent.  
	
	For the optimal theories with $c=3,4,5$ the symmetric B-matrices, which can be interpreted as graph adjacency matrix , describes the maximally connected graph
	\bea
	B_{ij}=\left\{\begin{array}{c}
		1,\quad i\neq j,\\
		0,\quad i=j. \end{array}
	\right.
	\eea
	Their enumerator polynomials and partition functions  can be found in  \cite{ds}.
	Here we only point out that for $p=2$ and $0\leq a,b<p$
	\bea
	\psi_{a,b}&=&{1\over |\eta|^2 }\sum_{n,m\in \Z} q^{(\tilde{a}+\tilde{b})^2/8}\bar q^{(\tilde{a}-\tilde{b})^2/8},\qquad \tilde{a}=a+2n,\quad \tilde{b}=b+2m,\\
	\psi_{0,0}&=& {|\theta_3(\tau)|^2+ |\theta_4(\tau)|^2\over 2  |\eta|^2},\\
	\psi_{1,1}&=&{|\theta_3(\tau)|^2-|\theta_4(\tau)|^2\over 2  |\eta|^2},\\
	\psi_{0,1}&=&\psi_{1,0}={|\theta_2(\tau)|^2\over 2 |\eta|^2}.
	\eea
	in full agreement with \cite{ds}.
	The spectral gaps are $\Delta^*=3/4,1,1$ for $c=3,4,5$ correspondingly. 
	
	\subsection{$c=6,7$}
	Optimal theories for $c=6,7$ were found in \cite{dymarsky2021non} to be related to codes, where a  construction, different from \cite{ds,dymarsky2021solutions}, relating codes over $F_4$ to CFTs was introduced. Here we show this construction is a particular case of a more general construction introduced in this work.
	
	Let us consider the glue matrix 
	\bea
	\label{Scase}
	\Lambda={1\over 3^{1/4}}\left(\begin{array}{cc}
		1 & -1\\
		-\sqrt{3} & -\sqrt{3}
	\end{array}\right).
	\eea
	This corresponds to $m=-1,n=1,k=0$ case as follows from 
	\bea
	g_\Lambda=\left(\begin{array}{cc}
		-2 & 0\\
		0 & 2
	\end{array}\right).
	\eea
	Clearly $\G=\Z_2 \times \Z_2$ which can be parametrized by $\ell^T=(a,b)$, $0\leq a,b<2$. As in the previous section we can identify $\G$ with $F_4$ via the Gray map
	\bea
	\label{F4}
	(a,b)\rightarrow c(a,b):=a\,\omega+b\,\bar\omega,
	\eea
	where $F_4=\{0,\omega,\bar \omega,1\}$.
	
	The scalar product inherited from on $\G$ from \eqref{ip}  reads
	\bea
	\eta((a_1,b_1),(a_2,b_2))=\ell_1^T g_\Lambda^{-1} \ell_2={b_1 b_2-a_1 a_2\over 2}.
	\eea
	Since the scalar product is defined up to integer shits, orthogonality with respect to $\eta$ is equivalent to orthogonality with respect to 
	\bea
	a_1 a_2+b_1 b_2\mod\, 2=c_1 c_2+\bar c_1 \bar c_2,  
	\eea
	where the right-hand-side uses notations \eqref{F4}. This is different from the conventional Hermitian scalar product on $F_4$
	\bea
	\label{F4scalar}
	(c_1,c_2)=c_1 \bar c_2+\bar c_1 c_2,
	\eea
	by an additional conjugation. 
	Thus a code $\C\in \G^c$ iso-dual with respect to scalar product on $\G$ inherited from \eqref{ip} and pairwise permutation $S$, $\C^\perp=S(\C)$, will be isodual to its conjugate, $\C^\perp=S(\bar \C)$, with respect to Hermitian scalar product \eqref{F4scalar}. This is exactly the isoduality condition outlined in \cite{dymarsky2021non}.
	
	Similarly, the evenness condition \eqref{evennesS}, written in coordinates
	\bea
	\sum_{i=1}^c {b_i b_{S(i)}-a_i a_{S(i)}\over 2} \mod 2=0,
	\eea
	matches precisely with the evenness condition of \cite{dymarsky2021non}.
	
	To complete the comparison with \cite{dymarsky2021non} we note that under Construction A \eqref{LambdaC} group elements $\ell^T=(a,b)$ will be mapped to 
	\bea
	v=\Lambda \ell,\quad \Lambda=\left(\begin{array}{cc}
		-{1\over 2} & -{1\over 2}\\
		{\sqrt{3}\over 2} & {-\sqrt{3}\over 2}
	\end{array}\right),
	\eea
	which is exactly the map from $c=a \omega+b \bar \omega\in F_4$ to  $\R^2$ used in 
	\cite{dymarsky2021non}. In other words, we have shown that the construction of  \cite{dymarsky2021non} is exactly the construction of this paper with the glue matrix taken to be \eqref{Scase}.
	
	We notice the choice $m=-1,n=1,k=0$ is not the canonical one. By an appropriate $GL(2,\Z)$ transformation we can bring it to $m=1, n=0, k=2$, satisfying \eqref{splagcondition2}. The new  form of the glue lattice generating matrix  is then 
	\bea
	\Lambda={2\over 3^{1/4}}\left(
	\begin{array}{cc}
		\frac{1}{2} & 1 \\
		\frac{\sqrt{3}}{2} & 0 \\
	\end{array}
	\right),
	\eea
	which is a hexagonal (triangular) lattice with the basic vector length $2/3^{1/4}$.
	
	The codes leading to optimal $c=6$ and $c=7$ theories, the hexacode and the ``septacode'' are rather bulky and we do not repeat them here. Let us just mention that in both cases  the  resulting spectral gap is $\Delta^*=\sqrt{4/3}$.

	\subsection{$c=8$}
	
	We consider the $n=m=0,k=4$ case with the glue lattice 
	\bea
	\Lambda=2\left(
	\begin{array}{cc}
		0 & 2^{1/4} \\
		2^{-1/4} & 0 \\
	\end{array}
	\right),\quad g_\Lambda=\left(\begin{array}{cc}
		0 & 4\\
		4 & 0
	\end{array}\right),\quad 
	\Lambda^\perp=
	\left(
	\begin{array}{cc}
		2^{1/4} & 0 \\
		0 & 2^{-1/4} \\
	\end{array}
	\right)/2.
	\eea
	In this case $\G=\Z_4\times \Z_4$ parametrized by ${\sf g}=(a,b)$, $0\leq a,b<4$ and
	$\ell^T=(a,b)$. Let us consider the code  $\C\in \G^8$ generated by rows of the following matrix
	\bea
	G_\C=\left(
	\begin{array}{cccccccccccccccc}
		0 & 0 & 0 & 2 & 2 & 0 & 0 & 0 & 0 & 0 & 0 & 0 & 0 & 0 & 0 & 0 \\
		0 & 0 & 0 & 0 & 2 & 2 & 0 & 0 & 0 & 0 & 0 & 0 & 0 & 0 & 0 & 0 \\
		0 & 0 & 0 & 0 & 0 & 2 & 2 & 0 & 0 & 0 & 0 & 0 & 0 & 0 & 0 & 0 \\
		1 & 1 & 1 & 1 & 1 & 1 & 1 & 1 & 0 & 0 & 0 & 0 & 0 & 0 & 0 & 0 \\
		2 & 0 & 0 & 0 & 2 & 0 & 0 & 0 & 0 & 0 & 0 & 0 & 0 & 0 & 0 & 0 \\
		3 & 1 & 0 & 0 & 3 & 1 & 0 & 0 & 2 & 2 & 0 & 0 & 0 & 0 & 0 & 0 \\
		0 & 3 & 1 & 0 & 0 & 3 & 1 & 0 & 0 & 2 & 2 & 0 & 0 & 0 & 0 & 0 \\
		0 & 0 & 3 & 1 & 0 & 0 & 3 & 1 & 0 & 0 & 2 & 2 & 0 & 0 & 0 & 0 \\
		3 & 0 & 0 & 3 & 1 & 0 & 0 & 3 & 0 & 0 & 0 & 2 & 2 & 0 & 0 & 0 \\
		1 & 3 & 0 & 0 & 3 & 1 & 0 & 0 & 0 & 0 & 0 & 0 & 2 & 2 & 0 & 0 \\
		0 & 1 & 3 & 0 & 0 & 3 & 1 & 0 & 0 & 0 & 0 & 0 & 0 & 2 & 2 & 0 \\
		0 & 0 & 0 & 0 & 1 & 1 & 1 & 1 & 1 & 1 & 1 & 1 & 1 & 1 & 1 & 1 \\
	\end{array}
	\right)
	\eea
	in the notations ${\sf c}=(a_1,\dots, a_8|b_1,\dots ,b_8)\in \C$. This code is even and self-dual with respect to 
	\bea
	\eta=\left(\begin{array}{cc}
		0 & I_8\\
		I_8 & 0
	\end{array}\right)/4.
	\eea
	Accordingly the lattice 
	\bea
	\Lambda_\C={1\over 2}\left(\begin{array}{cc}
		2^{1/4} I_8 & 0\\
		0 & 2^{-1/4} I_8
	\end{array}\right) (G_\C^T\, z   +4 k ),\quad z\in \Z^{12},\quad k \in \Z^{16},
	\eea
	is a Narain lattice with respect to 
	\bea
	g_L=\left(\begin{array}{cc}
		0 & I_8\\
		I_8 & 0
	\end{array}\right).
	\eea
	The lattice shortest vector has length $|v|^2=2\sqrt{2}$ yielding $\Delta^*=\sqrt{2}$.
	This follows from the lattice theta series, which can be readily obtained from the code enumerator polynomial. The code in question has $2^{16}$ codewords and enumerator polynomial $P_\C(x_{ab})$ is too large to be written explicitly here. Upon substituting $x_{ab}\rightarrow \psi_{ab}$ where 
	\bea
	\psi_{ab}=\sum_{k \in \Z^2} q^{|v|^2/2},\qquad v=\Lambda^\perp(\ell+4k),\quad \ell^T=(a,b).
	\eea
	(this definition is different from \eqref{psi} in two ways:  i) there is no $|\eta(\tau)|$ in the denominator because we are interested in the lattice theta-function rather than the CFT partition function ii) $\psi_{ab}$  depends on $q$ but not $\bar q$ as we are interested in the Euclidean structure only), we obtain 
	\bea
	P_\C(\psi_{ab})=1+4320 t^2+61440 t^3 +522720 t^4+2211840 t^5+O\left(t^{6}\right),\quad
	t=q^{2^{-1/2}}, \nonumber
	\eea
	which is exactly the theta-function of the Barnes-Wall lattice. This is in agreement with \cite{afkhami2020free}  who identified optimal $c=8$ theory to be based on a rescaled Barnes-Wall lattice, equipped with the Lorentzian metric and  understood as a Narain lattice.

	\section{Asymptotically large $c$}
	\label{largec}
	When $c\gg 1$ asymptotic behavior of spectral gap is not known.  Spinless modular bootstrap bounds $\Delta^*/c$ to be less than or equal to $1/\pi^2$ (with this value being obtained numerically) \cite{afkhami2020high}, while the full set of bootstrap constraints is likely to significantly decrease this value. Averaging over the whole moduli space of Narain theories provides a lower bound on $\Delta^*/c$ to be $1/(2\pi e)$ \cite{afkhami2020free}. Ref.~\cite{Dymarsky_2021}  conjectured this value to be asymptotically saturated, 
	\bea
	\lim_{c\rightarrow \infty} {\Delta^*\over c}={1\over 2\pi e}. \label{conjecture}
	\eea
	For this to be true, i.e.~for the mean value to (asymptotically) be the largest possible value, the distribution of spectral gaps on the Narain moduli space for large $c$ must be very sharply peaked around the mean without outliers. Thus, for consistency, as a necessary condition,  variance  should be very small. Using the ensemble of code CFTs, as well as chiral cousins of Narain theories,  ref.~\cite{Dymarsky_2021} has shown the variance of density of states distribution to be exponentially suppressed $\sim e^{-{\mathcal O}(c)}$, the conclusion consequently confirmed for the Narain theories in \cite{collier2021wormholes}. This does not constitute a proof of \eqref{conjecture} as 
	variance is not sensitive to possible outliers. 
	
	The conjecture of  \cite{Dymarsky_2021} is based on similarity between the ensemble of codes, ensemble of sphere packings, and the ensemble (space) of CFTs, and the problems of maximizing code Hamming distance, density of sphere packing and CFT spectral gap. Specifically for codes, there is an expectation that the Gilbert-Varshamov bound (the value resulting from averaging over all codes)  would asymptotically yield the best value of Hamming distance to code size ratio \cite{kozlov1969correcting,pierce1967limit}.  Similar expectation holds for the maximal density of lattice sphere packing: the densest packing to asymptotically saturate the Minkowski bound, which is simply averaged value over all possible lattices. (For sphere packings of general kind stronger asymptotic value is expected \cite{torquato2006new}.) While we leave validity of \eqref{conjecture} for future studies, here we show there are codes theories achieving this value of $\Delta^*$ for large $c$. 
	
	The Construction A used in this paper has a fundamental limitation: the corresponding lattices have vectors of certain length no matter how big the dimension $c$ is. This is formalized in equation \eqref{2dbound}, which provides an upper bound on $\Delta^*$. Thus, to obtain $\Delta^*$ scaling linearly with $c$ one has to adjust $n,m,k$ together with $c$ such that $|\G|$ grows as or faster than $c$. Here for simplicity we focus on the square gluing lattice $n=m=0$,  with prime $k=p$, discussed in sections \ref{sglc}, \ref{psglc}.  
	The spectral gap is given by \eqref{sgf} with $|v_\Lambda|=\sqrt{p}$,
	\bea
	\Delta^*={1\over 2}\min \left({d(\C)},p\right).
	\eea
	The best (maximal) generalized Hamming distance $d$ beyond small $c$ values is not known. One nevertheless can bound $d$ from below by consider ensemble averaging, the so-called Gilbert-Varshamov bound. Then, similarly to the case of binary linear codes one may expect best $d/c$ to asymptotically approach the bound when $c\rightarrow \infty$. 
	
	By averaged polynomial $\bar P(\{x_{ab}\})$ we mean  enumerator polynomial averaged over all $p^{c(c-1)/2}$ possible codes parametrized by $B$, see \eqref{Bcodes}. From the CFT point of view this is the calculation of averaged torus partition function.  
	So far we are interested only in $d$, or alternatively only in mass but not spin of the lightest non-trivial state, we can consider torus parameter to be purely imaginary $\tau=i\tau_2$. Then function \eqref{psglc} factorizes 
	\bea
	\psi_{ab}(i\tau_2)={1\over |\eta(\tau)|^2}\psi_a \psi_b,\qquad \psi_a=\Theta_{2a,p}(i\tau_2/2)=\sum_{k \in\Z}e^{-\pi \tau_2(a+ k p)^2/p}.
	\label{psia}
	\eea
	Going back to enumerator polynomial, instead of variables $x_{ab}$ we use 
	\bea
	x_{ab}=t_a t_b,\qquad t_a=t_{-a},
	\eea
	where the last property reflects $\psi_a=\psi_{-a}$. 
	We conjecture the form of corresponding averaged enumerator polynomial based on invariance under MacWilliams identity and explicit checks for sufficiently small $n$ and prime $p$, for which direct evaluation of $\bar P(\{t_at_b\})$  using computer algebra is feasible,
	\bea
	\nonumber
	\bar P(\{t_a t_b\})&=&{1\over p^{c(c-1)/2}} \sum_{B} P_{\C(B)}(\{t_a t_b\})=\\ &&t_0^{2c}+{\sum\limits_{k=0}^{p=1} \left(\sum\limits_{a=0}^{p-1}\sum\limits_{b=0}^{p-1} \cos\left({2\pi k a b\over p}\right) t_a t_b\right)^c-p\, t_0^c \left(\sum\limits_{a=0}^{p-1} t_a\right)^c\over p^c}.\label{barP}
	\eea
	
	Now we are ready to analyze this expression to deduce the lower bound on  $\Delta^*$. For large $c$, the main contribution to \eqref{barP} comes from $k=0$, leading to the averaged partition function
	\bea
\label{barZ}
	\bar Z\approx {1\over |\eta|^{2c}} {\left(\sum\limits_{a=0}^{p-1} \psi_a\right)^{2c}\over p^c}={1\over |\eta|^{2c}} {\left(\sum\limits_{n\in\mathbb Z} e^{-\pi \tau_2 n^2/p}\right)^{2c}\over p^c}.
	\eea
	Interpreted as sum over lattice points, the numerator is simply the sum over $2c$-dimensional square lattice of size $1/\sqrt{p}$. On the length scales of $\sim 1/\sqrt{p}$ or larger this is just the homogeneous distribution of points with the averaged density $1/(1/\sqrt{p}^{2c})=p^c$. This factor  exactly cancels $p^c$ in the denominator of \eqref{barZ} and we find density of states 
	\bea
	\rho(\Delta)={(2\pi)^c \Delta^{c-1}\over \Gamma(c)}
	\eea
	valid on scales $\Delta\gtrsim 1/\sqrt{p}$. This is exactly the density of states of ``$U(1)$-gravity'' -- Narain theory averaged  over the whole moduli space. Accordingly, the threshold for the density to become of order one is $\Delta=c/(2\pi e)$, which is our Gilbert-Varshamov bound. For this result to be valid we must require 
	$p/2>c/(2\pi e)$, otherwise shortest vector of $\Lambda_\C$ would have length $\sqrt p$. 
	
	To conclude, for sufficiently large $p$ we find that the averaged density of states (with zero chemical potential for spin) of $n=m=0$,
	and prime $k=p$ code theories is the same as the averaged density of states  for all Narain theories. In particular  in the limit $c\rightarrow \infty$, for $p>c/(\pi e)$ there are code theories with $\Delta^*/c=1/(2\pi e)$. Provided the conjecture of \cite{Dymarsky_2021} is correct, it would mean similar conjecture applies to $n=m=0$, prime $k=p$ codes, in the sense that their averaged Hamming distance is asymptotically the best one. 
	
	\section{Discussion}
	\label{conclusions}
	In this paper we proposed a family of constructions mapping additive codes over abelian groups $\G=\Z_p \times \Z_q$ to Narain lattices and hence Narain CFTs. Each construction is parametrized by a triple of integer numbers $n,m,k$ and an element from $O(1,1)$ parameterizing an even ``glue'' lattice $\Lambda\subset \R^{1,1}$.  The resulting Narain lattice $\Lambda_\C$ associated with a code $\C \subset \G^c$ obeys
	\bea
	\underbrace{\Lambda \oplus \dots \oplus \Lambda}_{c\, {\rm times}} \subset \Lambda_\C \subset \underbrace{\Lambda^\perp \oplus \dots\oplus \Lambda^\perp}_{c\, {\rm times}} \subset \R^{c,c}.
	\eea
	We call this glue construction following \cite{SPLAG}.  This construction generalizes and encompasses those  of \cite{ds,dymarsky2021solutions}, \cite{dymarsky2021non} and \cite{yahagi2022narain}. We call the CFTs obtained from codes ``code theories.'' Their torus partition functions $Z_\C$ are given in terms of the so-called code enumerator polynomials, which are multi-variable polynomials satisfying certain algebraic identities, which guarantee modular invariance of $Z_\C$. In this way one can construct many new solutions to modular bootstrap constraints. 
	
	We have shown that all conjectural optimal Narain theories for $c\leq 8$ identified in \cite{afkhami2021free}, meaning those with the largest spectral gap of $U(1)^c \times U(1)^c$ primaries are code theories. Furthermore we have shown there are code theories with the spectral gap $\Delta^*$ scaling linearly with $c\gg 1$, $\Delta^* \propto c/(2\pi e)$, with the coefficient which has been conjectured in \cite{Dymarsky_2021} to be maximal possible. The message of our work is clear: we conjecture that optimal Narain theories for any $c$ are code theories, either following from the constructions outlined in this paper, or their possible generalizations. 
	
	Speaking of the latter, one can straightforwardly generalize our construction by starting with a glue lattice $\Lambda \subset \R^{3,3}$ or in fact $\Lambda \subset \R^{r,r}$ for any $r\geq 2$. Another important direction would be to connect the bottom-up approach of this paper with the top-down approach of \cite{buican2021quantum} where quantum codes were given an interpretation in terms of CFT Hilbert space extended by defect operators. Finally, given our conjecture that optimal theories are code theories, it would be interesting to develop our approach into a practical way of constructing optimal theories with $c>8$, thus complementing conventional modular bootstrap. This would be an important but challenging task because there is no known efficient methods to construct ``good'' codes with largest or even large (generalized) Hamming distance. And though there is a finite number of codes for any given $\G$ and $c$, their number grows exponentially with $c$. Furthermore,  there is an infinite number of constructions, i.e.~infinite number of different $\G$ and $\Lambda$, making the problem seemingly  incomprehensible.  This pessimistic assessment could be too naive,  we expect only finite number of constructions to be relevant for any given $c$. The inequality \eqref{2dbound} as well as the results of section \ref{largec} clearly indicate $|\G|$ can not be too small, $|\G|^{1/2}\geq a c$, for $c\gg 1$ with some positive constant $a$. We also strongly suspect large generalized Hamming distance, for given $c$,  would require $|\G|$  not be too large. We conjecture this may come from the linear programming constraints stemming from the algebraic identities  satisfied by code enumerator polynomial (the MacWilliam identity and the condition due to code evenness), i.e.~generalization of Delsarte's bounds \cite{delsarte1973algebraic} to the types of codes of interest. For $c\gg 1$ we expect the bound  to have the form $|\G|^{1/2}\leq b  c$ with some $b>a$. Thus for large but finite $c$ we expect large but finite number of glue groups satisfying 
$b c \geq  |\G|^{1/2} \geq  ac$. This  form of the bound on $|\G|$ is merely a guess; the important point here is the expectation that  the problem of identifying the code with largest generalized Hamming distance can be reduced to an  optimization problem over a finite set. 
Of course even for moderate $c$  naive brute-force approaches such as going through all possible codes  very quickly becomes unfeasible. The resulting   optimization problem over a discrete set would be NP-hard, but novel quantum  platforms promise an exciting hope of solving medium-sized discrete optimization problems in real time, the avenue we hope to pursue in the future \cite{quantum}.
	
	To conclude, we would like to point out another very important direction for future work -- to extend the connection between codes and CFTs beyond the Narain theories.

	{\bf Acknowledgments.} The authors are 
	supported by the National Science Foundation under Grant No.~PHY 2013812. DC was also partially supported by the National Science Foundation under the Grant No. PHY 1818878. AD thanks Institut des Hautes $\acute{\rm E}$tudes Scientifiques, where this work was completed, for hospitality. 
	
	\appendix

	\section{Shortest vector bound}
	\label{Ebound}
	Let us consider a two-dimensional Euclidean lattice $\sf\Lambda$ with the scalar product $g_2$. Using rotation and up to an overall rescaling basis vectors  can be chosen to be $1$ and $\tau=\tau_1+i\tau_2$, where we introduced complex coordinates on $\R^2$. In other words
	\bea
	\label{g2}
	g_2={\sf \Lambda}^T {\sf \Lambda},\qquad {\sf \Lambda}=\alpha\left(\begin{array}{cc}
		1 & \tau_1 \\
		0 &\tau_2 \end{array}\right),
	\eea 
	and $\alpha$ is some scalar factor. Using $GL(2,\Z)$ transformations, together with an appropriate rotation and rescaling, we can bring $\tau$ to belong to fundamental domain 
	\bea
	|\tau_1|\leq 1/2,\quad \tau_2\geq 0,\quad |\tau|\geq 1. \label{fd}
	\eea
	In this case the shortest vector is $\alpha(1,0)^T$ and its norm is $\alpha^2$. 
	From \eqref{g2} we find $\alpha^4={\rm det}g_2/\tau_2^2$ and from \eqref{fd} we know $\tau_2^2\geq 3/4$. We therefore find the bound
	\bea
	\alpha^2\leq {2\over \sqrt{3}} \sqrt{{\rm det} g_2},
	\eea
	which in many cases is conservative. Applying that to \eqref{2dmetric} we obtain \eqref{2dbound}.  
	
	\bibliographystyle{JHEP}
	\bibliography{codes}
	
\end{document}